\documentclass{appolb}

\usepackage{graphicx}
\usepackage{subfigure}
\usepackage{cite}
\usepackage{epsfig}
\usepackage{amsmath,amsfonts}

\def\tr{{\rm tr}}

\def\MeV{\,\mathrm{MeV}}

\def\SU{\mathrm{SU}}

\newcommand{\ignore}[1]{}

\begin{document}

\title{Heavy Quark Entropy shift: From the Hadron Resonance Gas to
  Power Corrections\thanks{Talk by EM at {\em Excited QCD
      2016}, 6-12 March 2016, Costa da Caparica, Portugal.}
  \thanks{Work supported by the European Union (FP7-PEOPLE-2013-IEF)
    project PIEF-GA-2013-623006, Spanish Ministerio de Econom\'{\i}a y
    Competitividad and European FEDER funds (grant FIS2014-59386-P)
    and Junta de Andaluc\'{\i}a (grant FQM-225).  }  }
\author{E. Meg\'{\i}as\address{Max-Planck-Institut f\"ur Physik
    (Werner-Heisenberg-Institut), F\"ohringer Ring 6, D-80805 Munich,
    Germany} \and E. Ruiz Arriola, L.L. Salcedo \address{Departamento
    de F\'{\i}sica At\'omica, Molecular y Nuclear \\ and Instituto
    Carlos I de F\'{\i}sica Te\'orica y Computacional, \\ Universidad
    de Granada, E-18071 Granada, Spain} } \maketitle
\begin{abstract}

A heavy quark placed in the medium modifies its specific heat. Using a
renormalization group argument we show a low energy theorem in terms of the
defect in the trace of the energy-momentum tensor which allows the
unambiguous determination of the corresponding entropy shift after imposing
the third principle of thermodynamics for degenerate states.  We show how
recent lattice QCD data can be understood in the confined phase in terms
of a single-heavy hadronic spectrum and above the phase transition through power
corrections which are analyzed by means of a dimension 2 gluon condensate of
the dimensionally reduced theory.
\end{abstract}

\PACS{12.38.Lg, 11.30, 12.38.-t}
 
\bigskip
 
\section{Introduction}
\vbox
{
The Polyakov loop has been a major ingredient in the development of QCD at
finite temperature, where lattice calculations routinely used the bare
Polyakov loop as an order parameter for the crossover between
the confined hadronic phase and the quark-gluon plasma.  Its renormalization
was first accomplished in perturbative QCD (pQCD) by Gava and Jengo
\cite{Gava:1981qd}
}
\noindent
and non-perturbatively by the Bielefeld group \cite{Kaczmarek:2002mc}.  In a
series of works \cite{Megias:2005ve,Megias:2007pq,Megias:2009mp,Megias:2009ar}
we have pointed out that, instead of pQCD features, unexpected inverse power
corrections in temperature dominate above the phase transition, a
non-perturbative feature which finds a natural explanation in terms of
dimension-2 condensates of the dimensionally reduced theory.  On the other
extreme we have also found a hadronic representation in the confined phase
both for the Polyakov loop 
\cite{Megias:2012kb,Megias:2013xaa} 
and its correlators~\cite{Megias:2016onb} (see also~\cite{Megias:2015qya}). For a pedagogical review see e.g.~\cite{Arriola:2014bfa}.

Perturbative calculations of the renormalized Polyakov loop have been pursued to
NNLO~\cite{Burnier:2009bk,Brambilla:2010xn,Berwein:2015ayt}. The mysterious power
corrections have been confirmed on the lattice for $N_c=3,4,5$
\cite{Mykkanen:2012ri}.  Quite recently, the TUM lattice calculation
for physical quark masses in 2+1 flavours has been carried out in the
temperature range $125 \MeV \le T \le 6000 \MeV$~\cite{Bazavov:2016uvm} 
concluding that at earliest pQCD to NNLO~\cite{Berwein:2015ayt} might set in at
$5800 \MeV$.  Here we display the power correction pattern over a huge range
of temperatures and argue on the missing singly heavy hadronic states below
the phase transition.

\section{Renormalization Group}

The Polyakov loop is a local operator which in the static gauge reads $\tr
(\Omega (\vec x)) = \tr \left( e^{i g A_0(\vec x)/T} \right) $ ($ \tr ({\bf
  1})=N_c$). Its expectation value is a ratio of two partition functions (we
take conventionally $\vec x=0$)
\begin{eqnarray}
\langle \tr (\Omega) \rangle = \frac{Z_Q}{Z_0}=
\frac
{\int DA D\bar q D q  \, e^{-\int d^4 x \, {\cal L} (x)} \, \tr( \Omega)}
{\int DA  D \bar q D q \, e^{-\int d^4 x \, {\cal L} (x)}} \equiv e^{-\Delta F_Q/T} 
\end{eqnarray} 
where the QCD Lagrangian for $N_f=3$ flavours
$u,d,s$ reads, in terms of the re-scaled gluon field $\bar{A}_\mu = \sum_a
g A_\mu^a T_a $ with $\tr (T_a T_b) = \delta_{a b}/2 $,
\begin{eqnarray} 
{\cal L} (x) = -\frac{1}{4 g^2} 
(\bar G_{\mu \nu}^a)^2 + \sum_{q=u,d,s} \bar q (i \, {\slash \!\!\!\! D} - m_q ) q
\,.
\end{eqnarray} 
The renormalized Polyakov loop is uniquely defined up to a constant factor
which corresponds to $\Delta F_Q \to \Delta F_Q + c_\Omega$. There is no
natural way to fix the ambiguity, but it can be removed by using the
corresponding entropy
\begin{eqnarray} 
\Delta S_Q = - \frac{\partial \Delta F_Q}{\partial T}
= 
\frac{\partial}{\partial T} \left[ T \log \langle \tr (\Omega) \rangle \right] \, . 
\end{eqnarray} 
Being dimensionless, one should have $ \Delta S_Q = \varphi \left( g(\mu) ,
\log \frac{\mu}{T} , \log \frac{\mu}{m_q(\mu)} \right) $, with $\mu$ the
renormalization scale. Renormalization group invariance requires
\begin{eqnarray}
0= \mu \frac{d \Delta S_Q}{d\mu} \ = \beta(g) \frac{\partial \Delta S_Q}{\partial g} -
\sum_q  m_q (1+\gamma_q)  \frac{\partial \Delta S_Q}{\partial m_q} - T
\frac{\partial  \Delta S_Q}{\partial T}  
,
\end{eqnarray}
where the beta function and the mass anomalous dimension are, 
\begin{eqnarray} 
\beta(g) = \mu \frac{dg}{d\mu}\, ,  \qquad 
\gamma_q(g)= -\frac{d \log m_q}{d \log \mu} \, 
, 
\end{eqnarray}
respectively.  Direct evaluation yields the shift in the specific heat
\begin{eqnarray}
\Delta c_Q = T \frac{\partial  \Delta S_Q}{\partial T} 
= \frac{\partial}{\partial T}
\left\{ T \int d^4x  \left[ \frac{\langle \tr (\Omega)  \, \Theta(x) \rangle}{\langle \tr (\Omega) \rangle } - \langle \Theta(x) \rangle 
\right]  \right\}  \equiv \frac{\partial  \Delta U_Q}{\partial T}
,
\end{eqnarray}
where $\Theta$ is the trace of the energy momentum tensor,\footnote{Here we
  extend to $\Delta S_Q$ the argument of
  Ref.~\cite{Ellis:1998kj,Shushpanov:1998ce}.  The entropy shift is not a true
  entropy. For instance, the true entropy must be a monotonous function of the
  temperature since from $Z=\Tr e^{-H/T}$ it follows $c= T \partial_T S=
  \langle (H - \langle H\rangle)^2\rangle/T^2 >0$. Thus {\it both} $c_Q> 0$
  and $c_0> 0$ but the sign of $\Delta c_Q$ is not fixed.  The exact relations
  for thermodynamics of heavy quarks~\cite{Chernodub:2010sq} are still
  subjected to ambiguities which are removed in the specific heat.}
\begin{eqnarray}
\Theta \equiv \Theta^\mu_{\mu} = \frac{\beta (g)}{2g} \tr (G_{\mu \nu}^2) 
+ \sum_q m_q (1+\gamma_q) \bar q q  \,.
\end{eqnarray} 
The entropy shift can be obtained
by integrating with suitable boundary conditions featuring the
dimensions of the Hilbert space with and without Polyakov loop. At low
temperatures $Z_Q \sim 2 N_f e^{-M_0/T}$ 
for $N_f$ degenerate flavours
and $Z_0 \sim 1 $, whereas at
high temperatures $Z_Q \sim N_c Z_0$, thus
\begin{eqnarray}
\Delta S_Q (0) = \log (2 N_f) \, , \qquad  \Delta S_Q (\infty) = \log N_c 
\,.
\end{eqnarray} 
The first condition is the third principle of thermodynamics for degenerate
states. The recent TUM lattice calculations directly provide  the entropy in
the range $125 \MeV \le T \le 6000 \MeV$~\cite{Bazavov:2016uvm} taking the
convention $S_Q^{\rm TUM}(\infty)=0$ in harmony with their normalization
$e^{-F_Q^{\rm TUM}/T}= \langle \tr (\Omega) \rangle /N_c$ which, unlike ours,
{\it is not} a partition function at low temperatures. The critical
temperature was found to be $T_c=150 \MeV$.

\section{Singly Heavy Hadron Resonance Gas}

In the confined phase we expect a hadronic representation of the Polyakov loop
\cite{Megias:2012kb,Megias:2013xaa}. There the ambiguity comes from the heavy
quark mass which is subtracted from the hadron total mass. In
\cite{Megias:2012kb} we explicitly reconstructed the entropy $ d (T \log L(T))
/d T$ although the available lattice data were noisier than the recent
ones~\cite{Bazavov:2016uvm}.

\begin{figure}[tbc]
\begin{center}
\epsfig{figure=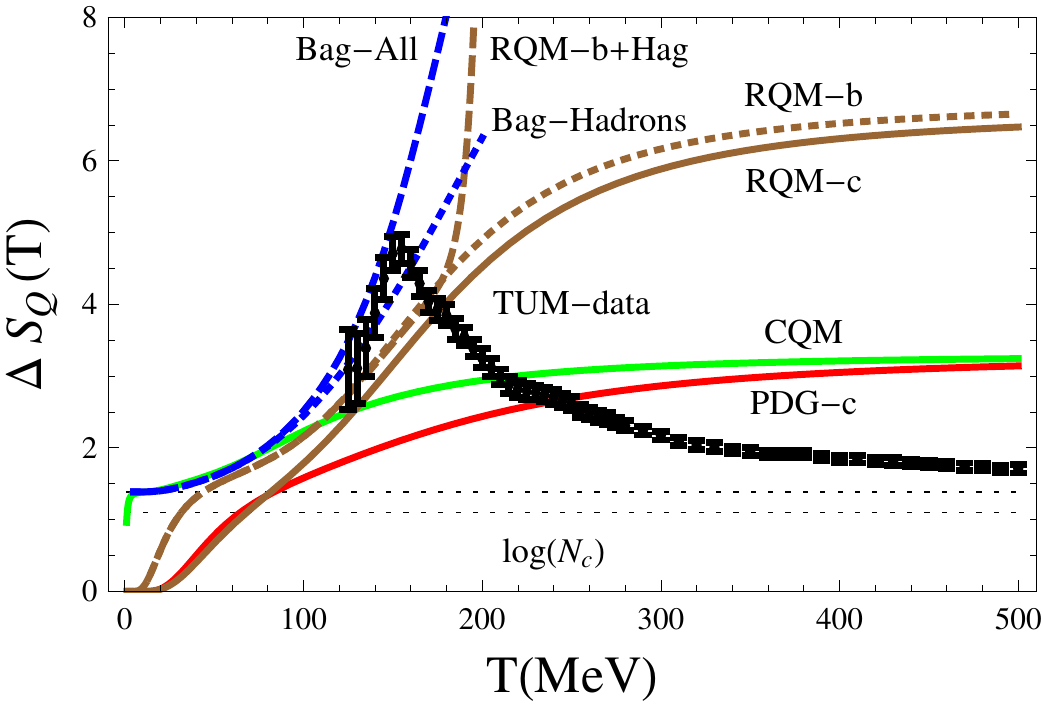,height=6cm,width=10.0cm}
\end{center}
\caption{The entropy as a function of the temperature.  We show
  results from various hadronic models: the bag model including all
  ($Q\bar q$, $Q qq$ and $Q\bar qg$) states and just hadrons, the RQM
   with one $c$- or $b$-quark and the PDG states with one
  $c$-quark. The Hagedorn extrapolation of the $b$-spectrum is also
  displayed. We also plot the CQM with $uds$ quarks and constituent
  mass $M=300 \MeV$ and the bare $m_u=2.5 \MeV $, $m_d= 5 \MeV$, $m_s=
  95 \MeV$ masses. Horizontal lines mark $\Delta S_Q(0) = \log 2 N_f$,
with $N_f=2$ the number of light degenerate flavours, 
  and $\Delta S_Q(\infty)= \log(N_c)$. Lattice data for 2+1
flavours are taken from
  Ref. \cite{Bazavov:2016uvm}.}
\label{fig:entropy-models}
\end{figure}

In Fig.~\ref{fig:entropy-models} we show the lattice entropy results
\cite{Bazavov:2016uvm} and compare them with the hadron resonance gas
using either the bag model (centered at the heavy quark source), the
PDG or the RQM of Isgur, Godfrey and Capstick for mesons and
baryons~\cite{Godfrey:1985xj,Capstick:1986bm} taking either the charm
or the bottom as the putative heavy quark. We have noted in previous
works that these RQM singly heavy states follow a Hagedorn-like
pattern with a Hagedorn-Polyakov temperature of about $T_H=200 \MeV$
for $b$-quarks.  Results from a simple constituent quark model (CQM)
are also shown
\begin{eqnarray} 
L = \sum_{q=u,d,s} g_q e^{-(M_{\bar Q q}-m_Q)/T} 
+ \sum_{q,q'=u,d,s} g_{q,q'} e^{-(M_{\bar Q q q'} - m_Q)/T} 
+ \cdots 
\end{eqnarray} 
with $M_{\bar Q q} = 2 M + m_q + m_Q$ and $M_{\bar Q q
  q'} = 3 M + m_q + m_{q'}  + m_Q$ and spin degeneracies $ g_q = 2 $, $
g_{qq'}= 4- \delta_{q,q'}$.

\begin{figure*}[tbc]
\begin{center}
\epsfig{figure=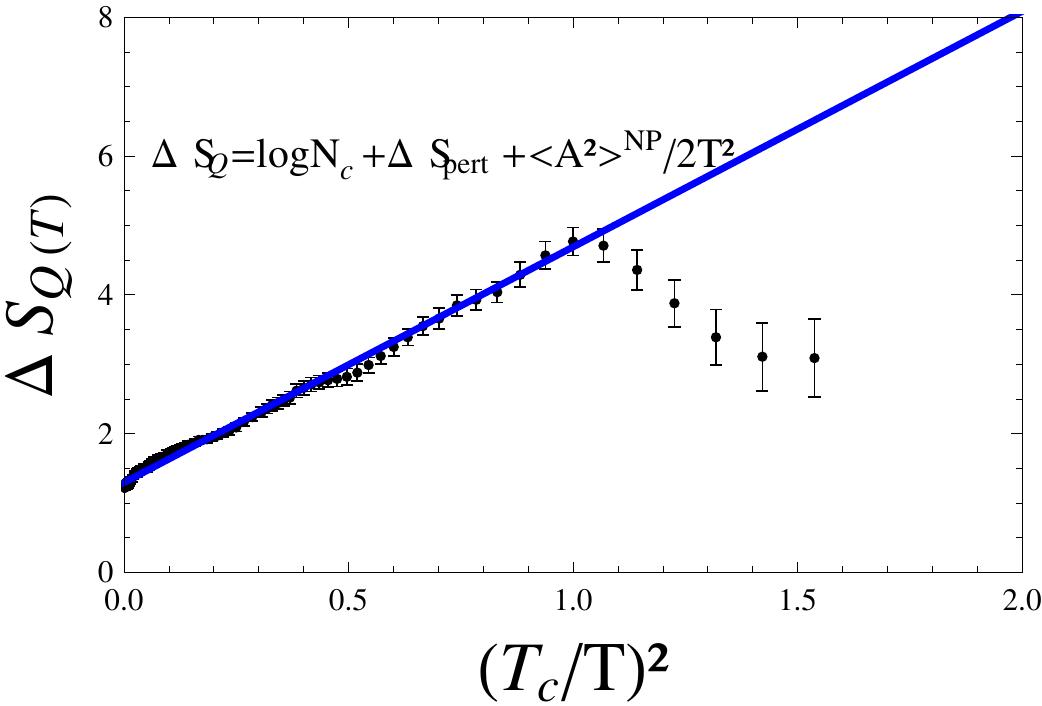,height=5cm,width=5.5cm}
\hskip.5cm 
\epsfig{figure=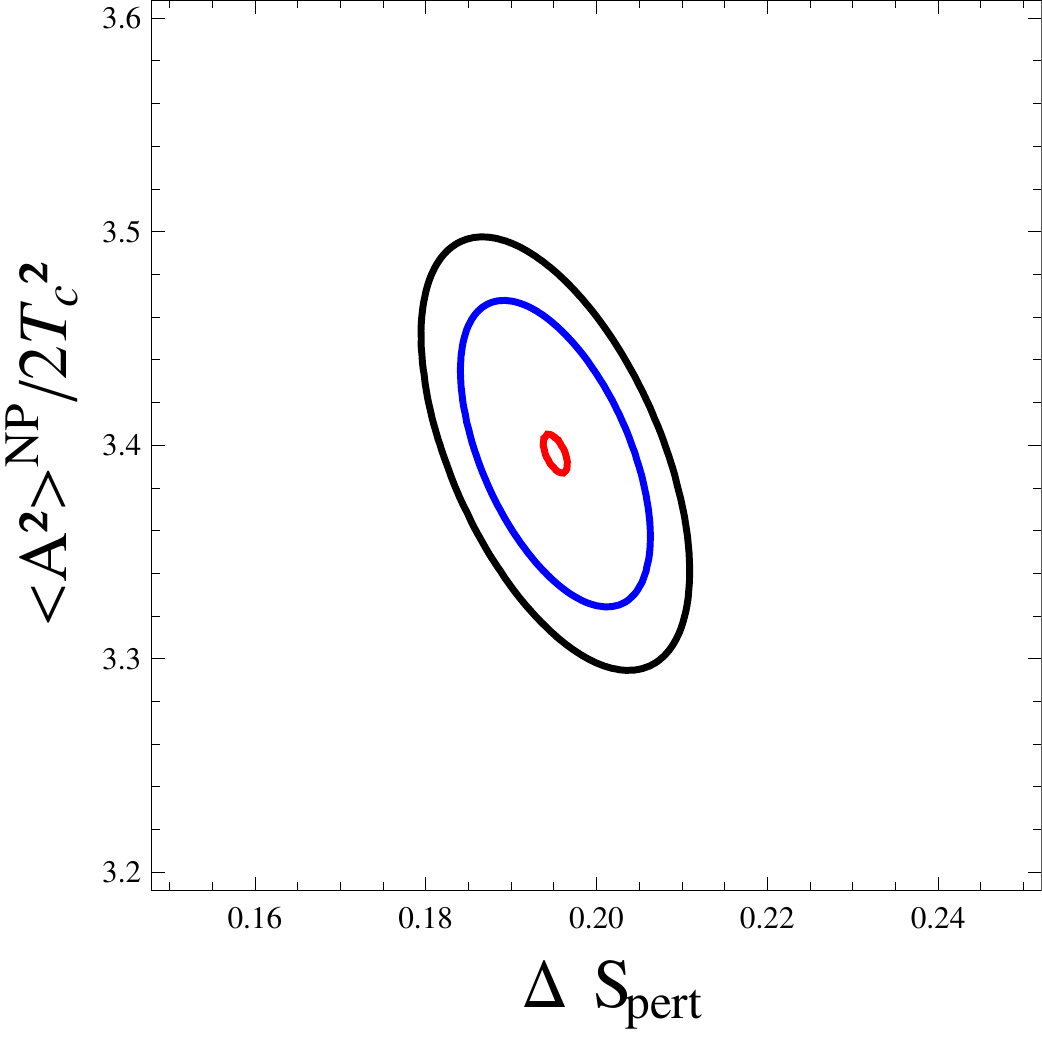,height=5cm,width=5.5cm}
\end{center}
\caption{Left panel: TUM lattice data for the entropy \cite{Bazavov:2016uvm}
  as a function of the inverse squared temperature in unites of the critical
  temperature. The straight line is the fit using the dim-2 condensate. Right
  panel: Correlation plot between the dim-2 condensate and the perturbative
  entropy.}
\label{fig:entropy-inv}
\end{figure*}

\section{Dimension-2 Condensate and power corrections}

At high temperatures the Polyakov loop can be expanded in powers of
$\bar{A}_0$ \cite{Megias:2005ve}
\begin{eqnarray}
\langle \tr (\Omega) \rangle = N_c - g^2 \frac{\langle (A_0^a)^2
  \rangle }{4 T^2} +O(g^6)
~ \sim ~ 
N_c \exp \left[- \frac{\langle \tr(\bar{A}_0^2) \rangle
  }{2 N_c T^2} \right] 
.
\end{eqnarray} 
$\langle (A_0^a)^2 \rangle$ has both perturbative (pert) and 
non-perturbative (NP) contributions, whence the entropy reads
\begin{eqnarray}
\Delta S_Q(T) = \log(N_c) + \Delta S_{\rm pert}(T) +
\frac{\langle A^2 \rangle^{\rm  NP}}{2 T^2}
,
\quad
\langle A^2 \rangle^{\rm  NP} \equiv
\frac{1}{N_c} \langle \tr (\bar A_0^2) \rangle^{\rm  NP}
,
\end{eqnarray}
where $\Delta S_{\rm pert}(T)$ is a slowly varying function with temperature.
In Fig.~\ref{fig:entropy-inv} we display the recent TUM data
\cite{Bazavov:2016uvm} as a function of $(T_c/T)^2$. A clear straight line
behaviour emerges over a huge range of temperatures. A fit for $T_c \le T \le
5000 \MeV$, neglecting the $T$ dependence in $\Delta S_{\rm pert}(T)$, gives
\begin{eqnarray}
\frac{\langle A^2 \rangle^{\rm NP}}{2 T_c^2} = 3.40(2) \,,  
\qquad  \Delta S_{\rm pert}(T) =0.195(3)
\,,
\end{eqnarray}
with $\chi^2=108$ for $N_{\rm dat}=92$. This gives $\chi^2/\nu=
108/(92-2)=1.2$ which is within the expected $1\pm \sqrt{2/\nu}$.

\section{Conclusions}

When a heavy colour source in the fundamental representation of the
$\SU(N_c)$ group is placed into the hadronic vacuum there arises an
entropy shift as a measurable and unambiguous observable. We noted
long ago that lattice data for the corresponding free energy display
corrections to the perturbative result in the unequivocal form of an
inverse second power of the temperature. This behaviour has been
corroborated by subsequent lattice studies. The present analysis
improves on previous ones thanks to the quality of the TUM lattice
data, and the fact that the unambiguous entropy is used for the
comparison. At low temperatures the quick increase in the number of
active states, $N = e^{\Delta S}$, suggests that there are missing
states in the singly heavy hadronic spectrum. At high temperatures
power corrections dominate over the perturbative contributions, in
harmony with our previous findings.  Taken at face value, the
impressive agreement calls for a deeper understanding on the nature of
these power corrections above the phase transition.



\end{document}